\documentstyle[aps,version2,preprint]{revtex}    
\tightenlines
\newcommand{\be}{\begin{equation}}
\newcommand{\ee}{\end{equation}}
\newcommand{\bea}{\begin{eqnarray}}
\newcommand{\eea}{\end{eqnarray}}
\begin{document} 

\begin{title}
Hall Conductance of a Pinned Vortex Lattice in a High Magnetic Field
\end{title}
\author{P.D. Sacramento}
\begin{instit}
Departamento de F\'{\i}sica and CFIF, Instituto
Superior T\'{e}cnico, \\ Av. Rovisco Pais, 1096 Lisboa Codex, Portugal
\end{instit}
\begin{abstract}
We calculate the quasiparticle contribution to the zero temperature
Hall conductance
of two-dimensional extreme type-II superconductors in a high
magnetic field, using the Landau basis.
As one enters the superconducting phase
the Hall conductance is renormalized to smaller values, with respect to 
the normal state result, until a quantum level-crossing
transition is reached. At high values of the order parameter,
where the quasiparticles are bound to the vortex cores, the Hall conductance
is expected to tend to zero due to a theorem of Thouless.
\end{abstract}

\section{Introduction}
Numerous superconductors like high-$T_c$ cuprates, A15's, boro-carbides and many
organics reveal the importance of Landau quantization at high magnetic fields and
low temperatures, as evidenced, for instance, by the recent observation of 
the de Haas-van Alphen oscillations
in the superconducting phase \cite{1}. In these systems the cyclotron 
splitting ($\omega_c$)
between the Landau levels (LL) is the largest energy scale, in particular
$\omega_c > \Delta$ (where $\Delta$ is the superconductor order parameter). 
The effect of the
Landau quantization has attracted theoretical attention and reveals itself by the
appearance of gapless superconductivity \cite{2,3} and power law dependence of 
the low-$T$
thermodynamic properties. A good description of the de Haas-van Alphen
oscillations has been obtained taking into account 
the Landau quantization \cite{4}. The gapless nature of the spectrum
also leads to an enhanced accoustic attenuation in contrast to usual
gapped superconductors \cite{bruun} and it survives even in the presence
of disorder \cite{s98}.

Close to the upper critical field, and at small values of $\Delta$, a diagonal
approximation (DA), where the coupling between the LL is neglected, is appropriate
\cite{3,bruun}. In this regime the quasiparticles propagate coherently throughout the vortex
lattice. Associated with the zeros of the order parameter in real space are gapless
points in the magnetic Brillouin zone. As $\Delta$ grows, the coupling between the
LL has to be taken into account. A perturbation scheme in the off-diagonal couplings
can then be carried out, as long as there are no level crossings \cite{5}, and it can be
shown analitically to all orders in the perturbation theory that there is 
always a discrete set of gapless points. In general, a numerical solution
of the Bogoliubov-de Gennes (BdG) equations has to be carried out. For large values
of $\Delta$ a regime is reached
where the quasiparticles are bound to the vortex cores with an energy spacing
typical of an isolated single vortex \cite{6}. In this regime one expects that a
tight-binding approximation should yield good results, since in this low-field regime
the vortex cores are sparse. Such a procedure has been carried out and a gapped regime
is indeed found \cite{6,7} signalling a transition from the high-field gapless regime
to the low-field gapped regime. 
Between the small $\Delta$ regime and this gapped regime 
several quantum level-crossings take place. This is a difficult numerical 
problem because the number of Landau levels increases considerably as the 
field lowers and it is not easy to determine
the value of $\Delta$ for which the transition to the gapped regime takes place. 

On the other hand, Thouless has shown that the Landau levels do not constitute a complete
basis to be used in a tight-binding approximation scheme and therefore it is
not possible to construct well localized representations of the magnetic
translation group
unless the Hall conductance, 
$\sigma_{xy}$,
is zero \cite{8}. 
Thouless showed that if Wannier functions are constructed they decay
with distance at most as $|\vec{r}-\vec{R}_i|^{-2}$ (if the Hall conductance
is nonvanishing) but he also showed that well localized Wannier functions
(decaying faster than any power) can be constructed if $\sigma_{xy}$ vanishes.
Using the Balian-Low theorem it has also been shown that in general \cite{Zak}
the Landau levels cannot be chosen sufficiently localized to make the
$\Delta x$ and $\Delta y$ uncertainties finite. This is equivalent to the
statement by Thouless on the slow decay rate of the Wannier functions
with distance. The notable exception is a sub-band with zero Hall conductance. 
Since it is intuitive that in the limit the vortex concentration 
is low
a tight-binding description should be appropriate, we expect, in the light of Thouless
theorem, that $\sigma_{xy}$ should be zero in this low-field regime. 
This is suggested by the localized nature evidenced by the numerical solution
of the BdG equations. To determine if the Wannier states are indeed a good basis
is a delicate and difficult matter and in general a tight-binding description
will have one conducting state at the band center \cite{etal}. We expect however
that in this low-field regime the functions will be well localized and
we argue therefore
that the Hall conductance could be used as an order parameter to signal the transition
from the high-field gapless regime, where Landau quantization has been 
shown to occur, (finite $\sigma_{xy}$) to the low-field gapped regime 
(zero $\sigma_{xy}$).

The calculation of the Hall conductance in the superconducting phase is also
interesting by itself. Several authors have paid attention to the Hall conductance
of the vortex lattice. In general, there are two contributions: one is due to the
vortex motion and the other to the quasiparticle contribution (usually associated with
modes localized in the normal region inside the vortex cores). One of the reasons
for this interest is that in the superconducting phase
$\sigma_{xy}$ has a different sign
with respect to the one of the normal phase \cite{9}. This has 
been shown to be due
to the vortex motion part since the localized modes are predicted to give a contribution
with the same sign as the normal phase value. 

In this paper we focus solely on the quasiparticle contribution 
which is the only remaining if the vortex lattice is pinned
to some imperfection (in this case the vortex motion is frozen). We do not address here
the sign change but study the influence of the coherent propagation of the
quasiparticles. Also, we take into consideration the Landau
quantization in a regime where a quasiclassical
approximation is not valid \cite{10}.

\section{Calculation of the Hall conductance}

We calculate the Hall conductance using the Kubo formula
\bea
\sigma_{xy}({\bf r}, {\bf r}^{\prime}) &=& -i \hbar L^2 \sum_{\beta \ne 0} \{
<0|J_x({\bf r}) |\beta> <\beta | J_y({\bf r}^{\prime})|0> \nonumber \\
&-& <0|J_y({\bf r}^{\prime}) |\beta> <\beta | J_x({\bf r})|0> \}
\frac{1}{(\epsilon_{\beta} -
\epsilon_0)^2} 
\eea
where the currents are given by
\bea
J_i({\bf r}) & = & \frac{e \hbar}{2 i mc} \sum_{\sigma} \{ \psi_{\sigma}^{\dagger}
({\bf r}) \left( \frac{\partial}{\partial x_i} \psi_{\sigma}({\bf r})\right) \nonumber \\
& - & \left( \frac{\partial}{\partial x_i} \psi_{\sigma}^{\dagger}({\bf r}) \right)
\psi_{\sigma}({\bf r}) \} 
- \frac{e^2}{mc^2} A_i \sum_{\sigma} \psi_{\sigma}^{\dagger} ({\bf r}) \psi_{\sigma}(
{\bf r})
\eea
Here $\sigma$ is the spin projection, $i=x,y,z$, ${\bf A}$ is the vector potential,
$\psi_{\sigma}({\bf r})$ is the electron field operator, the energies 
$\epsilon_{\beta}$ are the full solution of the many-body problem and $\beta=0$ is the
groundstate. We consider a square lattice of side $L$.

We calculate the Hall conductance obtaining
the energies from the solution of the BdG equations \cite{2,3}. The field
operators are written as
\bea
\psi_{\uparrow } ({\bf r}) & = & \sum_{\nu, {\bf q}} \left( u_{{\bf q}}^{\nu}({\bf r}) 
\gamma_{\nu, {\bf q}, \uparrow } - v_{{\bf q}}^{* \nu} ({\bf r}) 
\gamma_{\nu, {\bf q}, \downarrow }^{\dagger} \right) \nonumber \\
\psi_{\downarrow } ({\bf r}) & = & \sum_{\nu, {\bf q}} \left( u_{{\bf q}}^{\nu} ({\bf r}) 
\gamma_{\nu, {\bf q}, \downarrow } + v_{{\bf q}}^{* \nu} ({\bf r}) 
\gamma_{\nu, {\bf q}, \uparrow }^{\dagger} \right)
\eea
and
\be
\epsilon_{\beta} = \sum_{\sigma, \nu, {\bf q}} \epsilon_{{\bf q}}^{\nu} 
\gamma_{\nu, {\bf q}, \sigma}^{\dagger} \gamma_{\nu, {\bf q}, \sigma}.
\ee
Here $\gamma_{\nu, {\bf q}, \sigma}^{\dagger}$ creates a quasi-particle in the
level $\nu$, with momentum ${\bf q}$ and spin $\sigma$. The problem is diagonal
in ${\bf q}$ (vectors of the magnetic Brillouin zone) due to the periodicity
of the Abrikosov vortex lattice. The amplitudes $u$ and $v$ are expanded in the
Landau basis ($n$ is the Landau level) as
\be
u_{{\bf q}}^{\nu} ({\bf r})= \sum_n u_{{\bf q} n}^{\nu} \phi_{{\bf q} n} ({\bf r})
\ee
and
\be
v_{{\bf q}}^{* \nu} ({\bf r}) = \sum_n v_{{\bf q} n}^{* \nu} \phi_{-{\bf q} n} 
({\bf r}).
\ee
Here $\phi_{{\bf q} n} ({\bf r})$ are the eigenfunctions of the magnetic translation
group in the Landau gauge ($A_x=-Hy, A_y=0, A_z=0$) belonging to the $n^{th}$ Landau
level. The amplitudes $u_{{\bf q} n}^{\nu}$ and $v_{{\bf q} n}^{\nu}$ are the
solutions of the BdG equations and $\epsilon_{{\bf q}}^{\nu}$ are the energy
eigenvalues. 
The BdG equations to be solved are
\bea
\epsilon_n u_{{\bf q} n}^{\nu} + \sum_m \Delta_{nm} ({\bf q}) v_{{\bf q} m}^{\nu} &
= & \epsilon_{{\bf q}}^{\nu} u_{{\bf q} n}^{\nu} \nonumber \\
-\epsilon_n v_{{\bf q} n}^{\nu} + \sum_m \Delta_{n m}^{*}({\bf q}) u_{{\bf q} m}^{\nu} &
= & \epsilon_{{\bf q}}^{\nu} v_{{\bf q} n}^{\nu}
\eea
where $\epsilon_n = (n+1/2)-\mu/\hbar \omega_c$ ($\mu$ is the chemical potential),
$\Delta_{nm}({\bf q})$ is the matrix element of the order parameter $\Delta({\bf r})$
between electronic states (${\bf q}, n$) and ($-{\bf q}, m$) \cite{4}. 
In general, the order parameter is expanded in the LL for a charge $2e$, which leads
to a complicated structure for the BdG equations which have to be solved
self-consistently, since the order parameter satisfies $\Delta({\bf r})=
V < \psi_{\uparrow}({\bf r}) \psi_{\downarrow}({\bf r}) >$, where $V$ is the 
electron-electron interaction.
In the DA these equations are easilly solved
\bea
\epsilon_{{\bf q}}^{\nu} & = & \pm \sqrt{\epsilon_n^2 + |\Delta_{nn} ({\bf q})|^2} 
\equiv \epsilon_{{\bf q}}^{n}
\nonumber \\
u_{{\bf q} n} & = & \frac{1}{\sqrt{2}} \left( 1+ \frac{\epsilon_n}{\epsilon_{
{\bf q}}^{\nu}} \right)^{1/2} \nonumber \\
v_{{\bf q} n} & = & \frac{1}{\sqrt{2}} \left( 1- \frac{\epsilon_n}{\epsilon_{
{\bf q}}^{\nu}} \right)^{1/2}
\eea
(and therefore, $\nu \equiv n$). In this case
\be
u_{{\bf q} n}^{\nu} v_{-{\bf q} n}^{* \nu} = \frac{\Delta_{nn}({\bf q})}{2 \epsilon_{
{\bf q}}^{n}}.
\ee

As $\Delta$ grows, the off-diagonal terms 
become increasingly important. Recently, the leading corrections to the normal and
pairing self-energies have been found by delaying the Nambu rotation to the last step
and it was obtained \cite{5} that the renormalized quasiparticle energies can be written
as in the DA like 
\be
\epsilon_{n {\bf q}}^{\nu} = \pm \sqrt{ \bar{\epsilon}_n({\bf q})
+ |\bar{\Delta}_{nn}({\bf q})|^2}
\ee
but with renormalized normal and pairing terms which
to leading order are given by
\bea
\bar{\epsilon}_n ({\bf q}) & = & \epsilon_n + \sum_{p \neq 0} \frac{|\Delta_{n,n+p}
({\bf q})|^2}{p} \nonumber \\
\bar{\Delta}_{nn}({\bf q}) & = & \Delta_{nn}({\bf q}) \nonumber \\
& - & \sum_{p \neq 0, p^{\prime} \neq 0}
\frac{\Delta_{n,n+p}({\bf q}) \Delta_{n+p,n+p^{\prime}}^{*} (-{\bf q}) 
\Delta_{n+p^{\prime},n}
({\bf q})}{pp^{\prime}}
\eea
The effect of the off-diagonal couplings is taken into account by renormalizing the
diagonal (in the Landau index) terms keeping a diagonal problem to be
Nambu rotated ($\nu$ is still $\nu \equiv n$). 

In Fig. 1a we show the energy of the lowest band above the Fermi level,
obtained from eq. (8), as
a function of ${\bf q}$ (we only show one quadrant of the magnetic Brillouin zone
due to the square lattice symmetry) for $\Delta=0.1$ and $n_c=10$ (we fix
$\mu$ at this LL energy). In Figs. 1b and 1c we show the same band taking into account
the off-diagonal terms in eq. (7) considering $\Delta=0.1$ 
and $\Delta=0.5$, respectively. Results similar to these were previously
presented in other forms (see e.g. \cite{3,6}) but we present them here as well
to stress the role of the off-diagonal terms. 
The energy of the lowest band above the Fermi level is given by the gap
function $|\Delta_{nn}(\vec{q})|$ in the DA. This is given by \cite{3}
\be
\Delta_{nn}(\vec{q}) = \frac{\Delta}{\sqrt{2}} \frac{(-1)^n}{2^{2n} n!}
\sum_k e^{2 i k q_y a- (q_x+ \pi k/a)^2 l^2} 
H_{2n} [\sqrt{2} (q_x + \pi \frac{k}{a}) l]
\ee
where $a$ is the lattice constant of the square vortex lattice and $l$ is the
magnetic length, $l=\sqrt{\frac{\hbar c}{e H}}= \frac{a}{\sqrt{\pi}}$.
The gap $|\Delta_{nn}(\vec{q})|$ has zeros in the magnetic Brillouin zone
in momentum points ($q_j$) which are in direct correspondence with the positions
of the vortices in real space ($z_i$) such that $q_j l=z_i/l$ \cite{3}. These
gapless points are due to the center of mass motion of the Cooper pairs in the
presence of a high magnetic field and not due to the internal structure of the
gap function as in $d$-wave superconductors. The gapless points grow
considerably in number as $n_c$ grows \cite{3}. However, they preserve in the
magnetic cell the vorticity per unit cell of the order parameter
in real space. Away from the gapless points the gap raises to a scale of the order
of $\Delta$ giving rise to a complicated structure that oscillates ever more
strongly as $n$ grows. Increasing $\Delta$ only rescales the energy.
The energy has a linear dispersion around most of the zeros ($-1$ vorticity). Other
sets of zeros have a
quadratic dispersion for the square (triangular) lattice corresponding to a $-2$ ($-3$) 
vorticity, respectively. This dispersion leads to power law behavior at low
temperatures, as mentioned above.
Including
off-diagonal terms small gaps open in the spectrum \cite{5} which are due to the normal
(pairing) part of the self-energy and which are of order $\Delta^2$ ($\Delta^3$).
However, throughout the Brillouin zone the numerical prefactors are very
small and the spectrum is similar almost everywhere to the one obtained in the
DA. The notable exception are the Eilenberger points (EP) (e.g. ${\bf q}=(-0.5,0.5)$)
where the gap increases significantly as $\Delta$ grows \cite{5}. 
The off-diagonal matrix elements $\Delta_{nm}(\vec{q})$ have in general a different
structure of zeros. If $n+m$ is even the Eilenberger points remain but otherwise
they are absent.

These results are best summarized considering the density of states (DOS).
In Fig. 2 we plot the
DOS as a function of energy for different values of $\Delta$.
For small $\Delta$ the DOS is broadened from the Landau level locations by an amount
of the order of $\Delta$. The DOS is high at low energies. As $\Delta$ grows the energy
interval grows and eventually at $\Delta$ of the order of the cyclotron energy
the lowest band and the next band approach each other and a quantum level-crossing
transition occurs. Note that as $\Delta$ grows and the DOS spreads over to higher
energies, there remains a high DOS at low energies due to the low gap states
that retain the characteristic spectrum of the DA.
Therefore, it is expected that the DA will be appropriate at low $\Delta$ \cite{3,bruun}
(see however ref. \cite{6} for a thorough discussion).
 
To calculate the Hall conductance we
insert eqs. (2,3) in the expression for $\sigma_{xy}$ eq. (1) and 
obtain for the average Hall conductance 
\be
\sigma = \frac{1}{L^4} \int dx \int dx^{\prime} \int dy \int dy^{\prime} 
\sigma_{xy} (x,y;x^{\prime}, y^{\prime})
\ee
after several straightforward but lengthy steps ($q=2e$)
\bea
\frac{\sigma}{\frac{q^2}{h}} &=& -\sum_{\beta \neq 0} \frac{1}{(\epsilon_{\beta} -
\epsilon_0)^2} \sum_{\nu, \nu^{\prime}} \frac{1}{N_{\phi}} \sum_{n m} 
[ n_{\nu {\bf q} \uparrow} n_{\nu^{\prime} {\bf q} \downarrow} + 
 n_{\nu {\bf q} \downarrow} n_{\nu^{\prime} {\bf q} \uparrow}]^{\prime} \nonumber \\ 
& & \Re \{ [ v_{-{\bf q} n}^{\nu} u_{{\bf q} m}^{* \nu} ( u_{{\bf q} n+1}^{\nu^{\prime}} 
\sqrt{\frac{n+1}{2}} + u_{{\bf q}, n-1}^{\nu^{\prime}} \sqrt{\frac{n}{2}} ) 
( v_{-{\bf q} m+1}^{* \nu^{\prime}}
\sqrt{\frac{m+1}{2}} - v_{-{\bf q}, m-1}^{* \nu^{\prime}} \sqrt{\frac{m}{2}} ) ] 
\nonumber \\
& & +
[ v_{-{\bf q} n}^{\nu} u_{{\bf q} m}^{* \nu^{\prime}} ( u_{{\bf q} n+1}^{\nu^{\prime}}
\sqrt{\frac{n+1}{2}} + u_{{\bf q}, n-1}^{\nu^{\prime}} \sqrt{\frac{n}{2}} )
( v_{-{\bf q} m+1}^{* \nu}
\sqrt{\frac{m+1}{2}} - v_{-{\bf q}, m-1}^{* \nu} \sqrt{\frac{m}{2}} ) ] \}
\eea
The sum is extended over the excited states.
Since the current operator is a two-particle operator the excited states only differ
from the groundstate by the occupation of two quasi-particle (fermionic) states
(this is the reason for the prime in the number occupation term). Here $N_{\phi}$
is the number of momenta appearing in the sum which equals the number of vortices
in the system. The energies have been rescaled by $\hbar \omega_c $.

A selection rule in the DA implies that $\nu \neq \nu^{\prime}$. We can therefore
sum over the spin variables. The expression for $\sigma$ then takes the form
\bea
\frac{\sigma}{\frac{q^2}{h}} &=& -\frac{1}{4} \sum_{\beta \neq 0} 
\frac{1}{(\epsilon_{\beta} -
\epsilon_0)^2} \frac{1}{N_{\phi}} \sum_{{\bf q}} \sum_{n} \nonumber \\
& & \Re \{ (n_{n {\bf q}} n_{n+1, {\bf q}})^{\prime}  
(n+1) [\frac{\Delta_{nn}^{*}}{\epsilon_{{\bf q}}^{n}} 
\frac{\Delta_{n+1,n+1}}{\epsilon_{{\bf q} 
}^{n+1}} - 
\frac{\epsilon_{{\bf q}}^{n}-\epsilon_n}{\epsilon_{{\bf q}}^{n}} 
\frac{\epsilon_{{\bf q}}^{n+1}+
\epsilon_{n+1}}{\epsilon_{{\bf q}}^{n+1}} ] \nonumber \\
& & -(n_{n {\bf q}} n_{n-1, {\bf q}})^{\prime}  
(n) [\frac{\Delta_{nn}^{*}}{\epsilon_{{\bf q}}^{n}} 
\frac{\Delta_{n-1,n-1}}{\epsilon_{{\bf q}
}^{n-1}} -
\frac{\epsilon_{{\bf q}}^{n}-\epsilon_n}{\epsilon_{{\bf q}}^{n}} 
\frac{\epsilon_{{\bf q}}^{n-1}+
\epsilon_{n-1}}{\epsilon_{{\bf q}}^{n-1}} ] \}
\eea
Here $\epsilon_{\beta} = \epsilon_{{\bf q}}^{n} + \epsilon_{{\bf q}}^{n+1}$ in the first
term and $\epsilon_{\beta} = \epsilon_{{\bf q}}^{n} + \epsilon_{{\bf q}}^{n-1}$ in the 
second term. Fixing the chemical potential at the level $n_c$ ($n_c +1$ occupied levels)
and taking $\Delta=0$ (normal phase) we get that $\bar{\sigma} = \sigma h/q^2=n_c+1$,
as expected. The only contribution comes from the first term with $n=n_c$.
Taking now a small value of $\Delta$ the two dominant contributions come from the
previous term and from the first term with $n=n_c-1$. This leads to a discontinuity in
$\sigma$: at $\Delta=0$ the Hall conductance $\bar{\sigma}=(n_c+1)$ while at small,
but nonzero $\Delta$, $\bar{\sigma} \sim (n_c+1)-1/2$. As $\Delta$ grows $\bar{\sigma}$
decreases continuously. 

In Fig. 3 we show $\bar{\sigma}$ as a function of $\Delta$ for the DA and for
the leading order perturbation theory (PT) in the range
of values of $\Delta$ up to the order of the first level-crossing.
We consider the cases $n_c=4,10$.
As $\Delta$ grows the off-diagonal terms renormalize downwards the Hall conductance
with respect to the DA value. In the normal phase ($\Delta=0$) the Fermi level
(for a completely filled level) is in the gap between two Landau levels. As
$\Delta$ is turned on, $\mu$ is kept fixed at $\mu=n_c+1/2+\eta$ ($\eta \rightarrow 0$)
and the levels spread to higher energies which leads to a decrease of the Hall
conductance since the fraction of low energy states that may conduct decreases.
The presence of the off-diagonal terms fastens the rate of decrease for larger
values of $\Delta$, for the same reason. 
At very low $\Delta$ the two methods agree, as expected. 
Close to the
level crossing the study is very difficult because a fully self-consistent calculation
is needed to properly cover the crossover to the gapped regime. 
Going beyond the DA the several eigenfunctions of the BdG equations have
components in the various Landau indices, which due to the gapless (or
almost gapless) nature of the spectrum causes difficult numerical problems
due to the energy denominator in eq. (13). 
Also, strong mixing
of the LL destroys the LL structure beyond the level-crossing(s) transition(s). 
There is a discontinuity in the Fermi
level due to this level crossing (by $\pm 2\omega_c$ because of the doubling due to
the particle and hole bands, $u$ and $v$, respectively).
After the transition, $\bar{\sigma}$ is expeced to
decrease again and 
to tend to zero as $\Delta$ grows
even further (eventually after several level-crossings).  

\section{Conclusions}
 
In summary, we have calculated the Hall conductance
of a pinned vortex lattice in a high magnetic field using the Kubo
formula and the solution of the Bogoliubov-de Gennes
equations for the wave function amplitudes expanded in the Landau basis.
We compared the diagonal approximation with the leading order
perturbation theory recently introduced \cite{5}.
The Hall conductance decreases from the normal state value due
to the presence of the low-lying states immediately above the Fermi energy.
As $\Delta$ grows the spread in energy increases and $\sigma$ decreases.
We limited the study to the region of validity of the PT. We suggest that
$\sigma$ may be used as an order parameter to detect the transition from the gapless
regime (finite $\sigma$) to the gapped region (zero $\sigma$) where a tight-binding
description should be appropriate.


The author acknowledges discussions with Zlatko Te\v sanovi\' c at the early
stages of this work and for calling his attention to ref. 10. This work
was partially supported through PRAXIS Project /2/2.1/FIS/302/94.


\newpage

FIGURE CAPTIONS

Fig. 1- Lowest energy band as a function of ${\bf q}$ for a) $\Delta=0.1$
in the DA and including off-diagonal terms for b) $\Delta=0.1$ and c) $\Delta=0.5$,
respectively. 
Only one quadrant of the Brillouin zone is shown.

Fig. 2- Density of states (DOS) in arbitrary units as a function of energy for
$n_c=10$. The first two bands are shown.

Fig. 3- Hall conductance for $n_c=4,10$ as a function of $\Delta$ for the
two methods considered: diagonal approximation (DA) and leading order
perturbation theory (PT). The two points at $n_c+1$ are the $\Delta=0$ result
(normal phase).

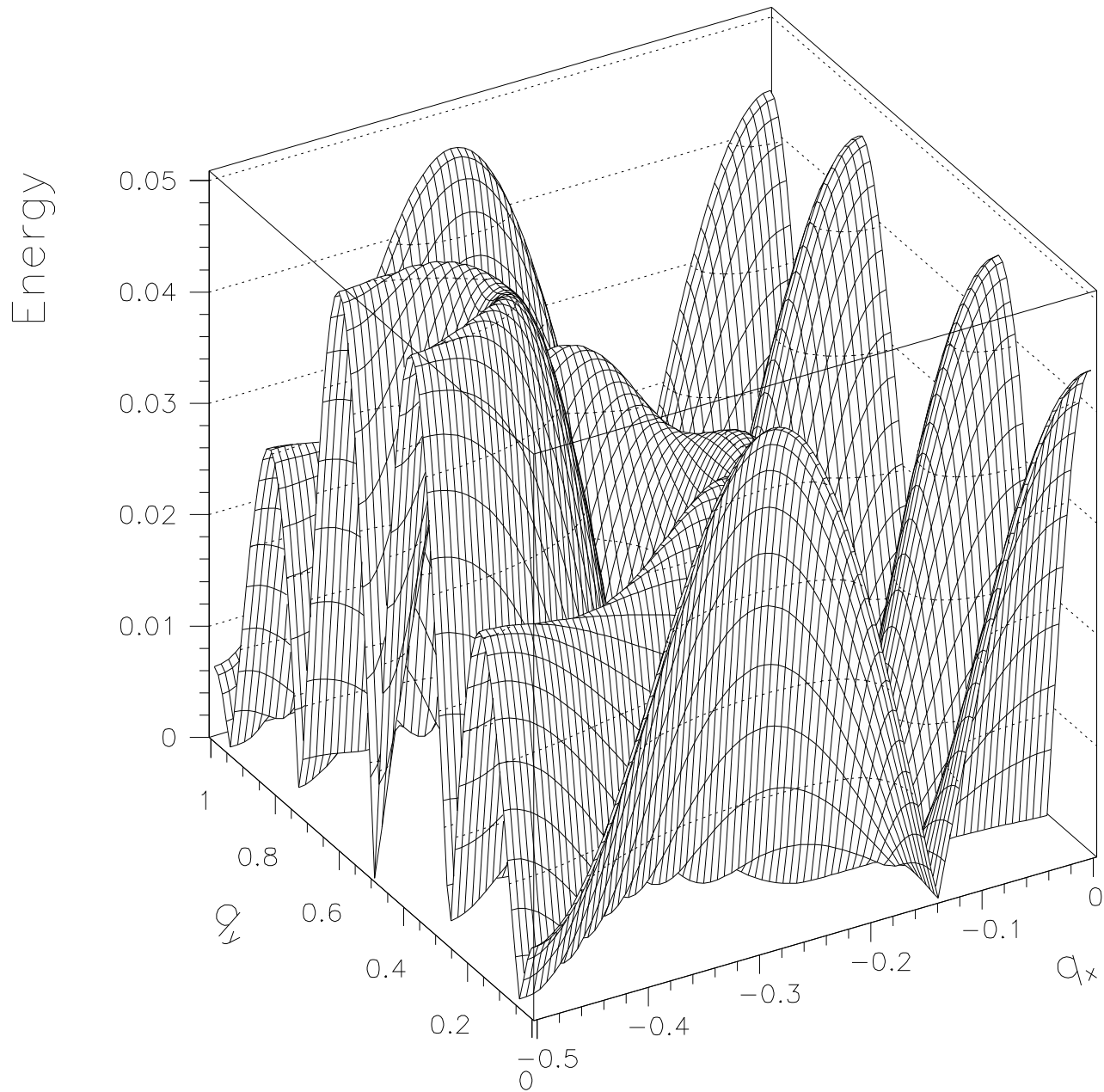
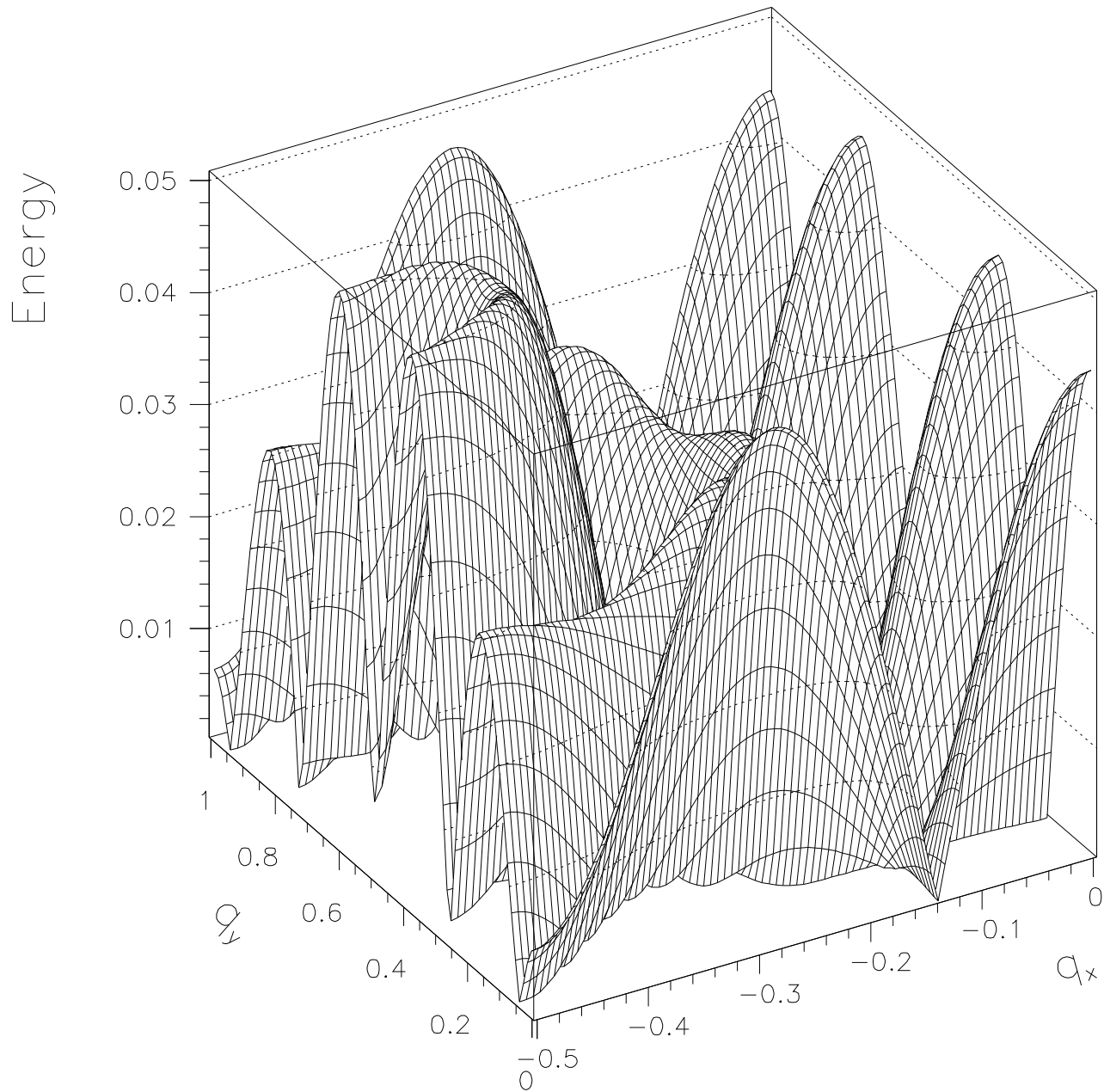
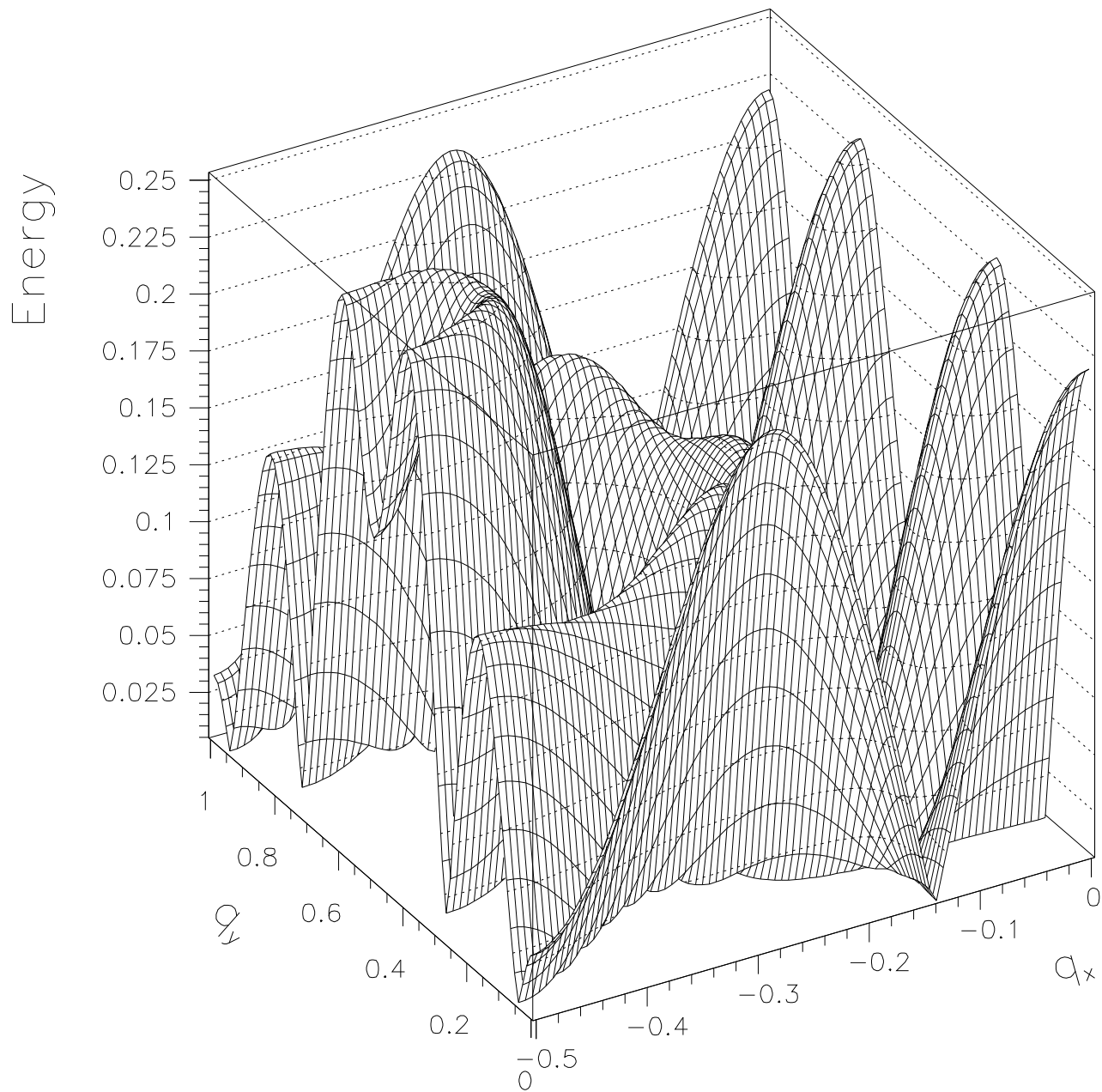
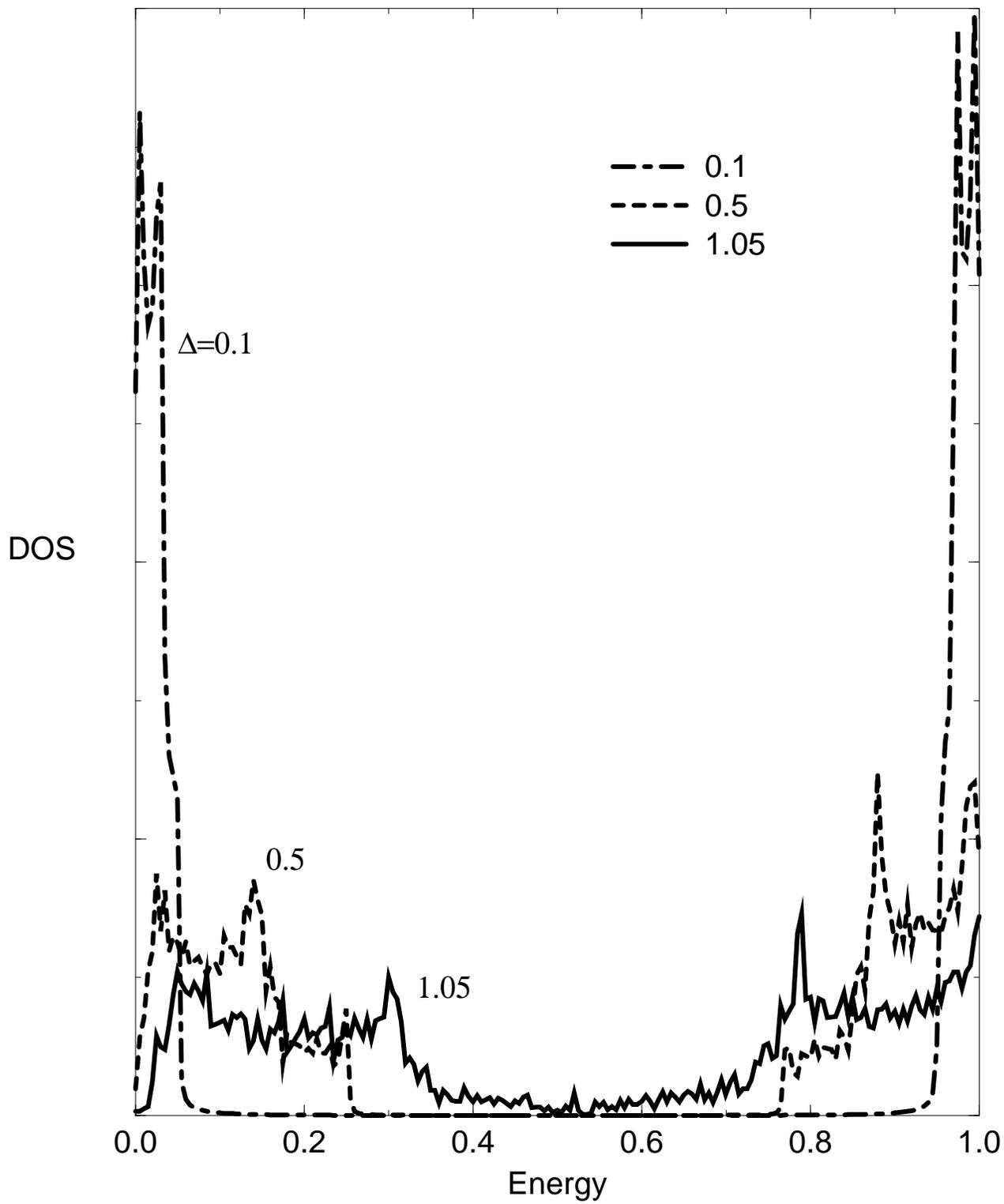
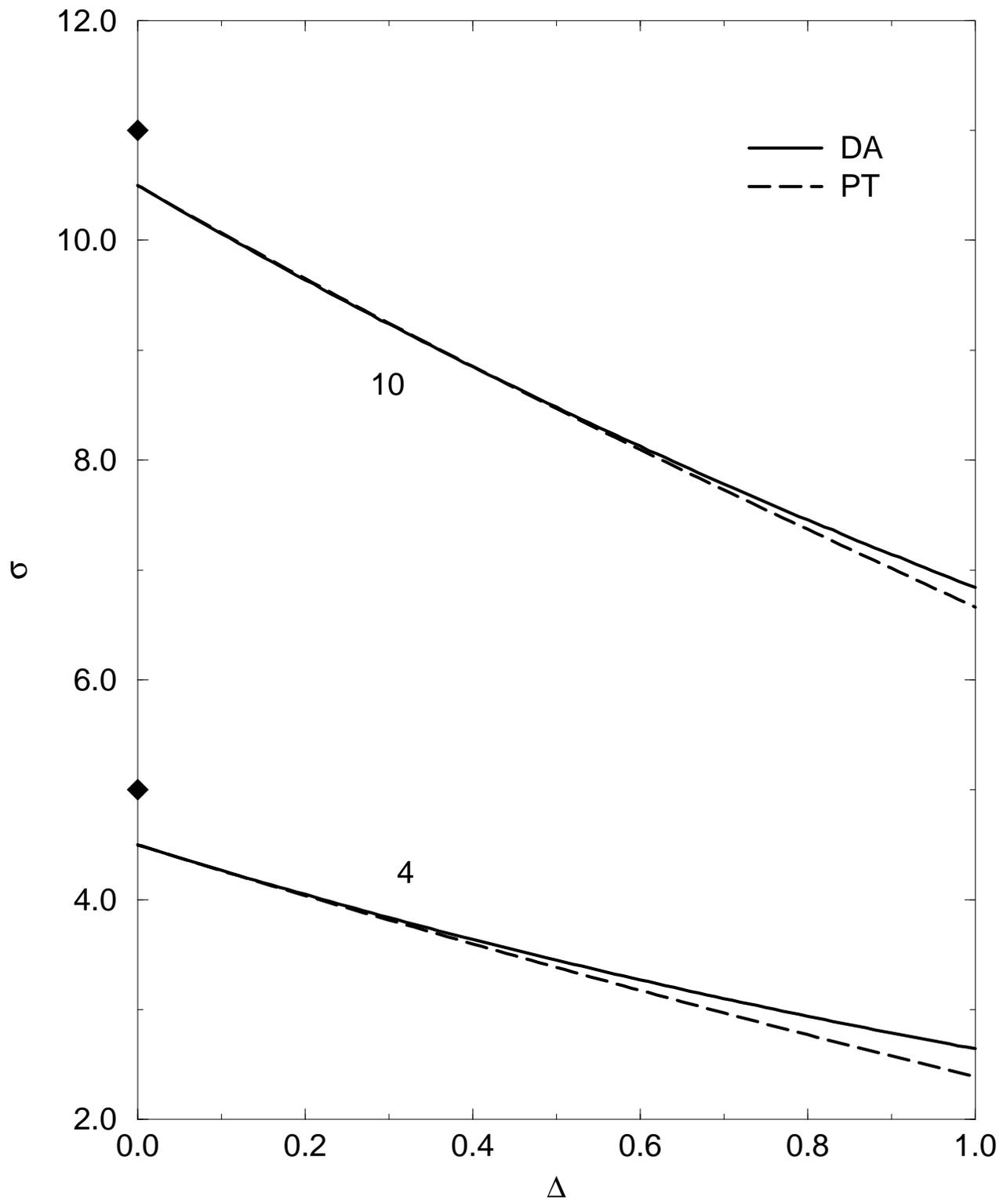
\end{document}